\documentclass{emulateapj}

\shorttitle{Compact IMBH X-ray binaries: potential LISA sources}
\shortauthors{Chen}

\begin{document}


\title{Compact intermediate-mass black hole X-ray binaries: potential LISA sources?}


\author{Wen-Cong Chen$^{1,2}$}
\affil{$^1$ School of Science, Qingdao University of Technology, Qingdao 266525, China;\\
 $^2$ School of Physics and Electrical Information, Shangqiu Normal University, Shangqiu 476000, China;\\
chenwc@pku.edu.cn
}



\begin{abstract}
The scientific aim of the space gravitational wave (GW) detector Laser Interferometer Space Antenna (LISA) that was scheduled to launch in the early 2030s is to detect the low-frequency GW signals in the Galaxy. Its main candidate GW sources are
compact binaries of white dwarfs and neutron stars. In this work, we examine whether compact intermediate-mass black hole (IMBH) X-ray binaries could be potential LISA sources. Our simulations indicate that IMBH binary with a 1000 $M_{\odot}$ IMBH and a 3 $M_{\odot}$ donor star in an orbit of initial orbital period near the so-called bifurcation period of 0.77 d could evolve into an ultra-compact X-ray binary, which will emit GW signals with maximum frequency of 2.5 mHz. According to the evolutionary tracks of characteristic strain, IMBH X-ray binaries with the initial donor-star masses of $1-3~M_{\odot}$ and the initial orbital periods slightly less than the bifurcation periods will be detectable by the LISA in a distance of 15 kpc. Assuming each of 60 Galactic globular clusters hosts a 1000 $M_{\odot}$ IMBH, the maximum number of compact IMBH X-ray binaries that LISA will detect in the Galaxy should be less than ten. Therefore, the detectability of compact IMBH X-ray binaries by the LISA is not optimistic.
\end{abstract}


\section{Introduction}
It marked the start of observational gravitational-wave (GW) astronomy that the first high-frequency GW signals from the double black hole merger event GW150914 in the distant galaxy \citep{abbo16} had been detected by the Advanced LIGO detectors \citep{aasi15}. Especially, the detection of the double neutron star coalescence GW170817 was a milestone in opening multi-messenger astronomy \citep{abbo17}.

A space GW detector Laser Interferometer Space Antenna (LISA) sponsored by the European Space Agency was scheduled to launch in the early 2030s \citep{amar17}. The scientific aim of the LISA will be to detect the low-frequency GW signals, and its sensitive frequencies are in a band between 0.1 mHz to 0.1 Hz, corresponding to orbital periods in the range of 20 s to 5 hours of binary systems \citep{sluy11}. In the Milky Way, there exist a number of classes of binary systems that the LISA would and should detect \citep{nele01,sluy11}. These potential LISA sources can be divided into the following two classes: the first one is detached binaries, including double white dwarfs, white dwarf - neutron star binaries, and double neutron stars; the second one is interacting binaries, i. e. CVs, AM CVn stars, and ultra-compact X-ray binaries.

Both population synthesis simulations and observation calibrations show that the expected number of double white dwarfs
can reach several hundred million in the Galaxy \citep{nele01,ruit10,liu10,yu10}. Based on semi-analytic model, some works
investigated the chirp of the emitting GW signals from double white dwarfs with the mass transfer \citep{hils00,kapl12,krem17}, and double neutron stars \citep{yu15}. Recently, \cite{taur18} performed a systematic work on the GW radiation of neutron star + white dwarf binaries, and found that ultra-compact X-ray binaries at a distance of 15 kpc can be detected by the LISA.

Recently, the existence of intermediate-mass black hole (IMBH) in globular clusters was supported by a growing number of observational and theoretical evidences \citep{port04a}. \cite{gair11} predicted that Einstein Telescope would detect thousands of IMBH binary mergers per year in the future. \cite{frag18} proposed that LISA and the Einstein
telescope will be best instruments detecting the merger events originating from the inspirals of IMBH-stellar mass black hole systems. \cite{mand08} suggested that the inspirals of stellar-mass compact objects into IMBHs in globular clusters can be detected by the advanced GW detectors. Most recently, \cite{jani20} showed that the inspiral, merger and ringdown of IMBH binaries will be detected by the LIGO and the LISA in multiband observations. However, the inspiral or merger of double IMBHs should be electromagnetically quiet. Compact IMBH X-ray binaries with stable mass transfer should be intriguing GW sources, which give us a chance to pursuit full multi-messenger investigations. The tidal capture or exchange encounters in young dense star clusters could result in that main sequence stars could spiral into and circularize around some IMBHs \citep{hopm04}. For example, the ultra-luminous X-ray source (ULX) in the young cluster MGG-11 of M82 may be an accreting IMBH from a captured donor star \citep{hopm04}. Because of the massive masses of IMBHs, IMBH binaries are possible to evolve into compact IMBH X-ray binaries via very strong gravitational radiation. \cite{port04b} found that IMBH X-ray binary with a donor star of $2~M_{\odot}$ and an initial orbital period of 0.5 days is visible as gravitational waves source in its whole lifetime. In this work, we attempt to examine whether compact IMBH X-ray binaries could be potential LISA sources.

\section{Evolution of compact IMBH X-ray binaries}
To obtain compact IMBH X-ray binaries, we calculate the evolution of some IMBH binaries.
The simulation employs a MESAbinary update
version (r12115) in the Modules for Experiments in Stellar
Astrophysics code (MESA; Paxton et al. 2015). The evolutionary beginning is
a binary system consisting of an IMBH (with a mass of
$M_{\rm bh}$) and a main sequence companion star (with a mass of $M_{\rm d}$)
in a circular orbit. The IMBH is assumed to be a point mass, and the chemical composition of the companion star is
X= 0.7, Y= 0.28, Z = 0.02. The loss of orbital-angular-momentum
plays a key role in influencing the evolution of IMBH binaries.
Three loss mechanisms of angular momentum by gravitational radiation,
magnetic braking \citep{rapp83}, and mass loss are included. If the donor star develops a convective envelope and possesses a radiative core, magnetic braking will turn on, and magnetic braking index $\gamma=3$. During the mass transfer, the accretion rate
onto the IMBH is limited to the Eddington accretion rate \citep{pods03}. The mass loss
from the vicinity of the IMBH is thought to carry away the specific
orbital angular momentum of the IMBH, while the donor
star winds carry away the specific orbital angular momentum of
the donor star. We run the MESA code until the donor star radius less than
the tidal radius ($R_{t}=(M_{\rm bh}/M_{\rm d})^{1/3}R_{\rm d}$, here $R_{\rm d}$ is the donor star radius) of the IMBH \citep{koch92}, or the time step less than the minimum time-step limit, or the stellar age greater than the Hubble timescale.

\begin{figure}
\centering
\includegraphics[width=1.15\linewidth,trim={0 0 0 0},clip]{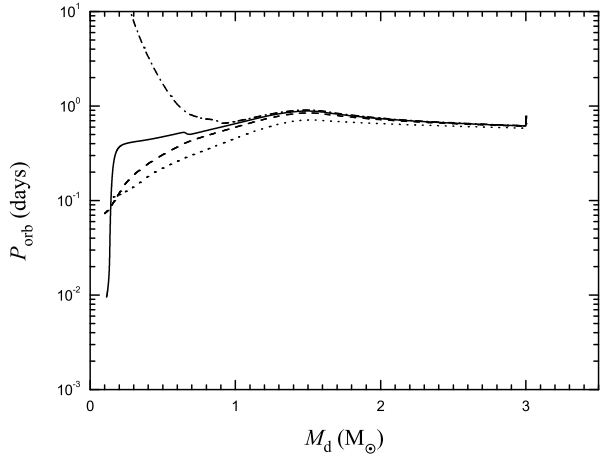}
\caption{Evolutionary tracks of IMBH binaries with a donor star mass of $3.0~\rm M_{\odot}$ and different initial orbital periods in the $P_{\rm orb} - M_{ \rm d}$ diagram. The solid, dashed, dotted, and dashed-dotted curves correspond to the initial orbital periods of 0.77, 0.75, 0.65, and 0.78 days, respectively.} \label{fig:orbmass}
\end{figure}

To examine whether can form compact IMBH X-ray binaries, we calculate the evolution of a large numbers of IMBH binaries with different initial orbital periods. The initial masses of the IMBHs and the donor stars are assumed to be 1000 $M_{\odot}$ and 3 $M_{\odot}$, respectively. In Figure 1, we plot the evolutionary tracks of four IMBH binaries in the $P_{\rm orb}-M_{\rm d}$ plane, where $P_{\rm orb}$ is the orbital period. It seems that there also exist a so-called bifurcation period like neutron-star low-mass X-ray binaries \citep{sluy05a,sluy05b}. In principle, the bifurcation period is related to the angular momentum loss mechanisms \citep{pyly88,pyly89,ergm98,pods02,ma09}. Based on the angular momentum loss mechanisms adopting in this work, the bifurcation period of IMBH binary with a donor star of $M_{\rm d}=3~M_{\odot}$ is 0.77 d. Our simulated results are same to the conclusion given by \cite{chen16}, IMBH binaries with initial periods near the bifurcation period tend to form more compact IMBH X-ray binaries. For example, when the initial period is $P_{\rm orb,i}=0.77$ d, the IMBH binary will evolve into a compact IMBH X-ray binary with orbital period $P_{\rm orb}=14$ min; when $P_{\rm orb,i}=0.75$, and 0.65 d, the minimum orbital periods will be 1.7 and 2.6 hours. During the mass transfer, the angular momentum loss is dominated by the gravitational radiation, which is at least 1-2 orders of magnitude higher than the magnetic braking. For an initial orbital period of 0.78 d greater than the bifurcation period, the IMBH binary will evolve into an IMBH X-ray binary with wide orbit (see also Figure 1). It is possible to evolve into compact ($P_{\rm orb}\la 2~\rm hours$) IMBH X-ray binaries for IMBH binaries with initial orbital periods less than the bifurcation period.

Figure 2 shows that the evolutionary track of the X-ray luminosity for IMBH X-ray binary with $P_{\rm orb,i}=0.77$ d. We compute the X-ray luminosity by $L_{X}=0.1\dot{M}_{\rm bh}c^{2}$, here $\dot{M}_{\rm bh}$ is the accretion rate, $c$ is the light velocity in vacuo. After 34.5 Myr nuclear evolution, the donor star overflows its Roche lobe, and transfers its H-rich material to the IMBH. In the most life, the IMBH X-ray binary appear as a normal X-ray source, and the X-ray luminosity is in the range of $\sim10^{36}$ to $10^{37}~\rm ergs\, s^{-1}$. At the final stage, the X-ray luminosity sharply increases due to the rapid shrinkage of the orbit. When the age of the donor star is 3672.65 Myr, the X-ray luminosity exceeds $10^{39}~\rm ergs\, s^{-1}$, and the IMBH X-ray binary becomes an ultra-luminous X-ray source (ULX). Such a system only last $32000$ yr in the ULX stage, hence ULXs with compact orbits are very rare in observations.

\begin{figure}
\centering
\includegraphics[width=1.15\linewidth,trim={0 0 0 0},clip]{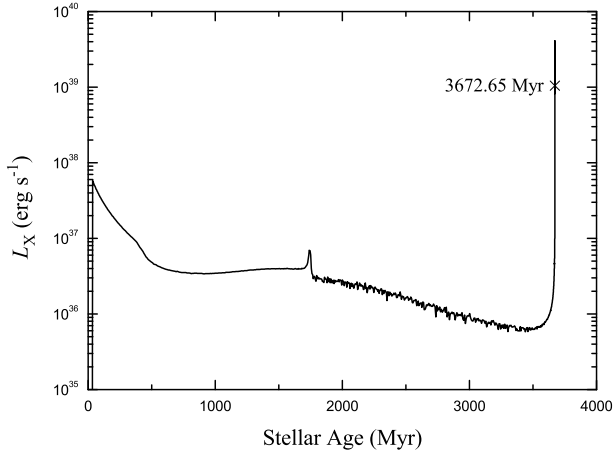}
\caption{Calculated X-ray luminosity as a function of the stellar age for IMBH
X-ray binary with $M_{\rm bh}=1000~M_{\odot}$, $M_{\rm d}=3~M_{\odot}$, and $P_{\rm orb,i}=0.77$ d. The cross represents the moment that IMBH X-ray binary appears as ULX.} \label{fig:orbmass}
\end{figure}

\section{Detectability of LISA}
\begin{figure}
\centering
\includegraphics[width=1.15\linewidth,trim={0 0 0 0},clip]{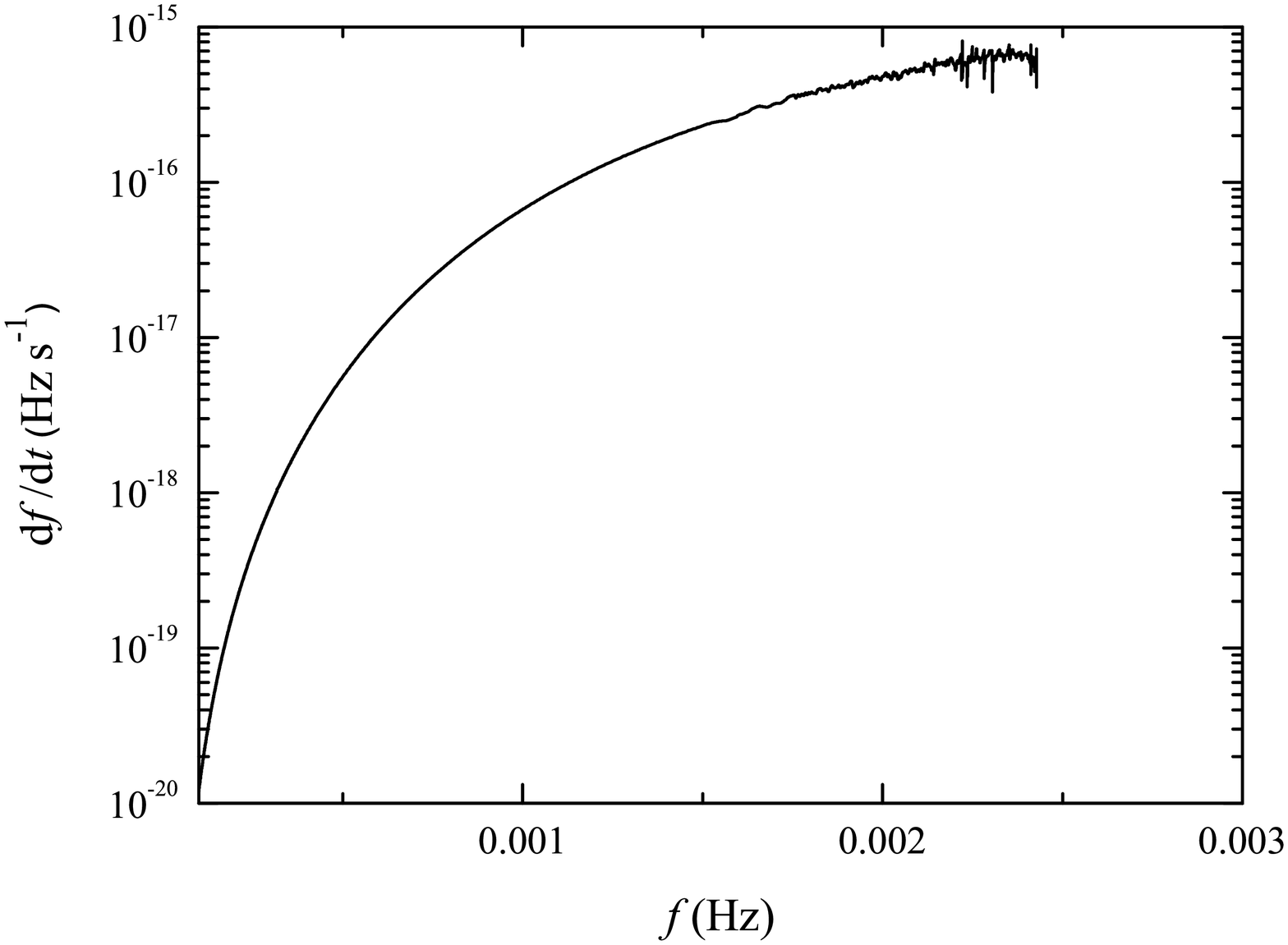}
\caption{ Evolutionary track of the frequency derivative $\dot{f}$ of GW signals in the $\dot{f}~-~f$ diagram
for the evolutionary final stage of the IMBH X-ray binary plotted in Figure 2.} \label{fig:orbmass}
\end{figure}

Figure 3 shows the evolution of the frequency derivative $\dot{f}$ as a function of the frequency $f$ ($f=2/P_{\rm orb}$) of GW signals for the evolutionary final stage of the IMBH X-ray binary plotted in Figure 2. It is clear that the maximum of $\dot{f}$ is close to the maximum of $f$. When $f_{\rm max}=2.5$ mHz, the maximum frequency derivative $\dot{f}_{\rm max}\approx7.0\times10^{-16}~\rm Hz\,s^{-1}$. For a detection duration $T=4~\rm yr$, the maximum frequency change is $\Delta f_{\rm max}=9.0\times10^{-8}$ Hz. Therefore, the GW signals emitting by such an IMBH X-ray binary can be thought as being monochromatic. By accumulating the power in the signal over cycles of $Tf$ for an observation duration $T$, the characteristic strain of GW can be written as \citep{finn00,taur18}
\begin{equation}
h_{\rm c}\approx\sqrt{2fT}h_{0},
\end{equation}
where
\begin{equation}
h_{0}=\left(\frac{32}{5}\right)^{1/2}\frac{\pi^{2/3}G^{5/3}f^{2/3}\mathcal{M}^{5/3}}{c^{4}d_{L}},
\end{equation}
is the GW amplitude producing by a compact binary at luminosity distance $d_{L}$ \citep{evan87}, where $G$ is the gravitational constant. Inserting typical parameters, and taking $T=4~\rm yr$, we have
\begin{equation}
h_{\rm c}\approx 2.5\times 10^{-20}\left(\frac{f}{1~\rm mHz}\right)^{7/6}\left(\frac{\mathcal{M}}{1~M_{\odot}}\right)^{5/3}\left(\frac{15~\rm kpc}{d_{L}}\right),
\end{equation}
and the chirp mass
\begin{equation}
\mathcal{M}=\frac{c^{3}}{G}\left(\frac{5\pi^{-8/3}}{96}f^{-11/3}\dot{f}\right)^{3/5}.
\end{equation}

An accurate measurement for the frequency derivative $\dot{f}$ by the LISA requires a very large signal-to-noise ratio (SNR). The resolution on the frequency derivative $\dot{f}$ can be estimated to be \citep{taur18}
\begin{equation}
\dot{f}_{\rm min}\approx 2.5\times 10^{-18}\left(\frac{100}{SNR}\right)\left(\frac{4~\rm yr}{T}\right)^{2}~\rm Hz\,s^{-1}.
\end{equation}
For the IMBH X-ray binary in Figure 3, $\dot{f}=2.5\times 10^{-18}~\rm Hz\,s^{-1}$ when $f=0.4$ mHz. This implies that the LISA will be able to detect the chirp signals from such a system when $f>0.4~ \rm mHz$ for a SNR above 100.

\begin{figure*}
\centering
\includegraphics[width=2.0\columnwidth]{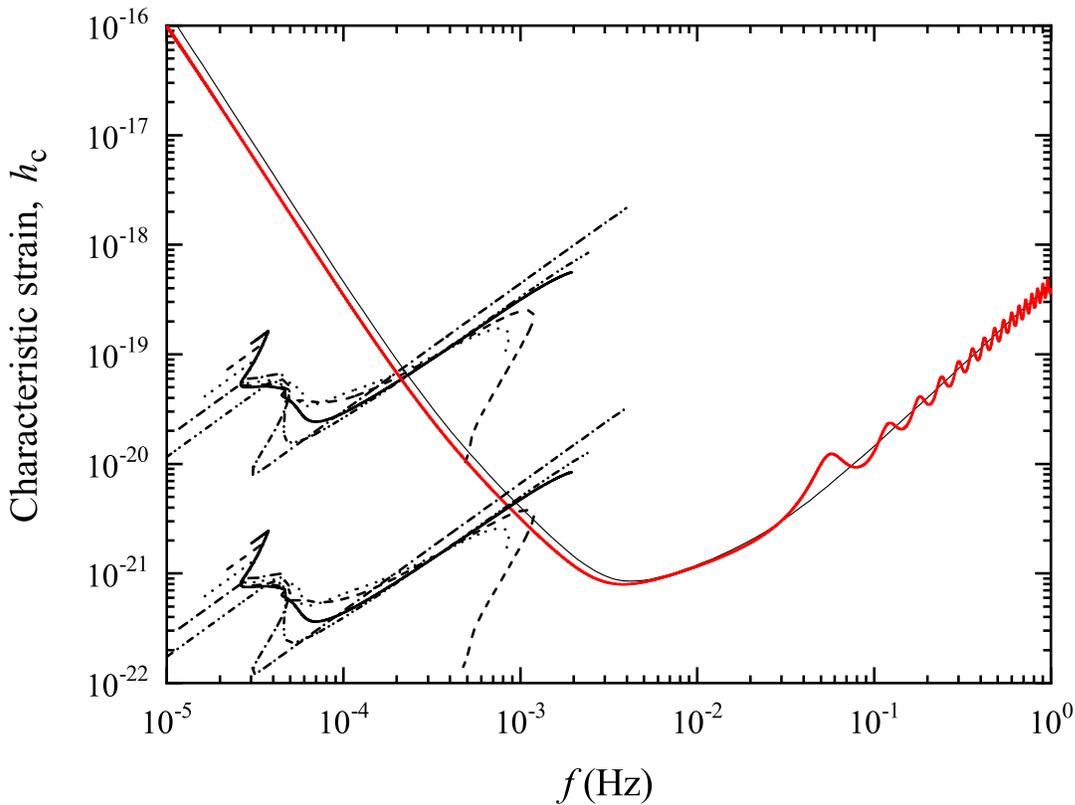}
\caption{Evolutionary tracks of four IMBH X-ray binaries in the characteristic strain amplitude vs GW frequency diagram. The red curve is the LISA sensitivity curve based on 4 yr of observations. The solid, dashed, dotted, dashed-dotted, and dashed-dotted-dotted curves correspond to the evolutionary tracks of IMBH binaries with $M_{\rm bh}=1000~M_{\odot}$, and $M_{\rm d}=3.0,2.5,2.0,1.5$, and $1.0~M_{\odot}$, respectively. All initial orbital periods of IMBH binaries are the bifurcation periods. The upper and under curve groups represent the luminosity distance of 15 kpc and 1 Mpc, respectively.} \label{fig:orbmass}
\end{figure*}


Figure 4 shows the evolutionary tracks of five IMBH X-ray binaries in the characteristic strain amplitude vs GW frequency diagram.
The LISA sensitivity curve by the black curve originates from a good analytic model (see also equation 13 in Robson et al. 2018), while the red curve is the results given by the numerical calculation. When $d_{L}=15$ kpc, five IMBH X-ray binaries with an initial orbital period equalling to the bifurcation period can penetrate the LISA sensitivity curve, and appear as potential LISA sources. Even if $d_{L}=1$ Mpc, IMBH X-ray binaries with donor star masses of 3.0, 1.5, and $1.0~M_{\odot}$ are still visible as the LISA sources.

To obtain the initial parameter space of the progenitors of IMBH X-ray binaries that can be detected by the LISA, we have simulated the evolution of a great number of IMBH binaries in the $P_{\rm orb,i}-M_{\rm d,i}$ plane. In Figure 5, the IMBH binaries are denoted by the filled circles if their descendants will be visible by the LISA for $d_{L}=15$ kpc. The solid curve shows the bifurcation periods of IMBH binaries for different donor-star masses. All systems above this curve will evolve into IMBH X-ray binaries with long orbital period (including the cases $M_{\rm d}=3.5~M_{\odot}$, and $P_{\rm orb}\geq0.7~\rm d$), which cannot enter the sensitive frequency scope of the LISA. The bifurcation periods display a tendency, in which a larger donor-star mass correspond to a smaller bifurcation period. It is clear that the progenitors of LISA sources tend to have an orbital period slightly lower than the bifurcation period. IMBH binaries with initial orbital periods obviously lower than the bifurcation period are difficult to evolve into compact orbits that the LISA can detect.

In Table 1, we summarize some evolutionary quantities of IMBH binaries with their respective bifurcation periods for different initial donor-star masses. All IMBH binaries with initial orbital periods equalling to the bifurcation periods can evolve into ultra-compact IMBH X-ray binaries with orbital periods less than 34 min (which corresponds to a GW frequency of 1 mHz). The initial GW frequency that these source are visible by the LISA is 0.22-0.23 mHz. The timescales of IMBH X-ray binaries as LISA sources are roughly 7-54 Myr, while they appear as luminous X-ray source ($L_{\rm X}>10^{38}~\rm erg\,s^{-1}$) only last $\sim0.02-2.0~\rm Myr$. In principle, compact IMBH X-ray binaries are difficult to become ULXs. When $M_{\rm d}=2.0$, and $2.5~M_{\odot}$, the donor stars evolve into plant-like mass objects that are absent in surface hydrogen content, just like the donor star of black widow pulsar PSR J1311-3430 \citep{plet12}. It was already noticed that the initial orbital periods should be less than and near the bifurcation period in order to form a donor star like PSR J1311-3430 \citep{benv13}.

We next estimate the maximum number of compact IMBH X-ray binaries that LISA will detect. It is generally thought that IMBHs reside in globular clusters, young dense clusters, and dwarf galaxies. For young dense clusters, it is too young to produce such compact IMBH X-ray binaries. In the Galaxy, there exist $137$ globular clusters. However, \cite{holl08} found that only 60 Galactic globular clusters are able to retain IMBHs with an initial mass of $1000~M_{\odot}$. Assuming the number of IMBHs $N_{\rm IMBH}=60$, the number of IMBH X-ray binaries appearing as LISA sources can be estimated as
\begin{equation}
N_{\rm IMBH, LISA}=N_{\rm IMBH}\Gamma P t_{\rm GC}t_{\rm LISA}/t_{\rm ev}.
\end{equation}
According to the simulation given by \cite{hopm04}, the rate that IMBHs capture companions is $\Gamma\sim5\times 10^{-8}$. Assuming that initial companion masses ($1-3~M_{\odot}$) and initial orbital periods ($0.5-3.5$ days) obey a uniform distribution, the probability that the IMBH binaries satisfies the parameter space of the progenitors of potential LISA sources $P\sim0.1$ should be an optimistic estimation. The age $t_{\rm GC}$ of the globular clusters is approximately two times as long as the evolutionary timescale $t_{\rm ev}$ of IMBH X-ray binaries. The timescale that compact IMBH X-ray binaries are visible as LISA sources $t_{\rm LISA}\sim 10~\rm Myr$. Therefore, the maximum number that IMBH X-ray binaries appear as LISA source in the Galaxy should be less than ten. This number is obviously smaller than those of ultra-compact X-ray binaries ($\ga 100$, Tauris 2018), AM CVn, and double white dwarf systems \citep{nele01}.

\begin{table*}
\begin{center}
\caption{Selected Evolutionary Properties for IMBH binaries with their respective bifurcation periods for different initial donor star masses. \label{tbl-2}}
\begin{tabular}{@{}lllllllllll@{}}
\hline\hline\noalign{\smallskip}
$M_{\rm d,i}$ & $P_{\rm bif}$ &  $P_{\rm rlov}$ &$X_{\rm c,rlov}$ &$X_{\rm sur,min}$ &$P_{\rm min}$ & $M_{\rm d, min}$ &$\bigtriangleup t_{\rm LISA}$& $f_{\rm i,LISA}$ & $\bigtriangleup t_{\rm ULX}$& $\bigtriangleup t_{\rm LX}$\\
 ($ M_{\odot}$)     &  (d)  & (d)   &   &    &    (min) & ($ M_{\odot}$ )    & (Myr)& (mHz) & (Myr)& (Myr)\\
\hline\noalign{\smallskip}
1.0 & 3.23 &0.60&1.8e-11&0.17& 14& 0.12 & 8.4& 0.23 & 0.02&0.13\\
1.5 & 1.97 &0.79&0.19& 0.90&11&0.13 &6.9 & 0.22 & 0.006&0.02\\
2.0 & 1.41 &0.73&0.43&3.9e-5& 34 &0.009 &19.2&0.22 & 0&1.8\\
2.5 & 1.05 &0.66&0.56&2.6e-13&21&0.0016 &54.4&0.23 &0 &1.9\\
3.0 & 0.77 &0.62&0.65&0.07&14& 0.11& 9.6& 0.23 &0.03 &0.3\\
\hline\noalign{\smallskip}
\end{tabular}
\tablenotetext{}{}\\{Note. The columns list (in order): the initial donor-star mass, bifurcation period, orbital period, central H abundance at the beginning of Roche lobe overflow, final surface H abundance, minimum orbital period, final donor star mass, duration that the IMBH X-ray binary is visible as LISA source, initial GW frequency appearing as LISA source, duration that the IMBH X-ray binary appear as ULX, and duration that the IMBH X-ray binary appear as luminous source ($L_{\rm X}>10^{38}~\rm erg\,s^{-1}$).}
\end{center}
\end{table*}

\begin{figure}
\centering
\includegraphics[width=1.15\linewidth,trim={0 0 0 0},clip]{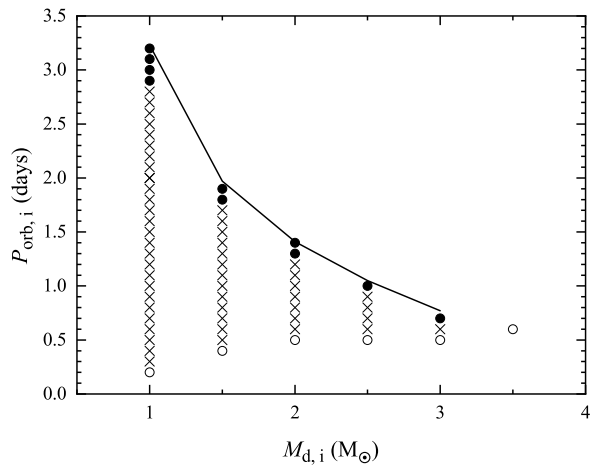}
\caption{Parameter space distribution of the progenitors of IMBH X-ray binaries
that can be detected by the LISA in the initial orbital period vs. initial donor-star mass diagram.
The initial masses of IMBHs are assumed to be $1000~M_{\odot}$, and the luminosity distance is $d_{L}=15~\rm kpc$. The solid curve represents the bifurcation periods of IMBH binaries with different donor star
masses. The crosses represent the IMBH binaries that can evolve into compact IMBH X-ray binaries, while they cannot enter the sensitivity scope of characteristic strain of the LISA. The open circles denote IMBH binaries that the donor stars have already filled their Roche lobe at the beginning of binary evolution.} \label{fig:orbmass}
\end{figure}

\section{Discussion and Summary}
Detectability of inspirals of stellar-mass compact objects into IMBHs \citep{mand08} or double IMBHs \citep{amar06,veit15}, and merger of IMBHs and compact objects \citep{mape10,hast16} or double IMBHs \citep{mats04,abad12,mazz14} as GW sources were extensively studied. However, compact IMBH X-ray binaries with stable mass transfer present a chance to pursuit full multi-messenger investigation, providing many valuable information of stellar and binary evolution. In this work, we diagnose whether IMBH X-ray binaries can be intriguing GW sources that the LISA will detect. Employing the MESA code, we simulate the evolution of a great number of IMBH binaries, and find that the systems with initial orbital periods lower than the bifurcation period can evolve into tight orbits due to the strong gravitational radiation. When $M_{\rm bh}=1000~M_{\odot}$, $M_{\rm d}=1.0-3.0~M_{\odot}$, all IMBH binaries with initial orbital periods equalling to the bifurcation periods can evolve into compact IMBH X-ray binaries, which can easily be detected by the LISA at a distance of $d_{L}=15~\rm kpc$. Especially, several compact IMBH X-ray binaries will still be visible as LISA sources even if it is at a distance of $d_{L}=1~\rm Mpc$ (see also Figure 4).

We also explore parameter space of the progenitors of IMBH X-ray binaries as potential LISA sources. For initial donor star masses in the range of $1.0-3.0~M_{\odot}$, and initial orbital periods in the range of 0.7-3.2 days (depending on the donor star masses), the corresponding IMBH binaries can evolve into promising LISA sources. Some IMBH binaries with shorter orbital periods can evolve into compact IMBH X-ray binaries, however, they will not be detectable for the LISA due to the low characteristic strain. After inspirals of normal stars into IMBHs and circulation, orbital angular momentum conservation predicted an orbital separation $a\sim(4-5)(M_{\rm bh}/M_{\rm d})^{1/3}R_{\rm d}$ (Hopman et al. 2004). This indicates that the initial orbital periods of IMBH binaries are approximately 2-4 days \citep{li04}. Therefore, the parameter space that we predict is reasonable, and the relevant IMBH binaries could form via tidal \citep{hopm04} or dynamical capture (Baumgardt et al. 2004, however, dynamical capture events tend to produce IMBH binaries with orbital periods of 100-1000 d) in globular clusters. Similar to the conclusion of \cite{port04b}, IMBH X-ray binaries with massive donor stars of $\ga4~M_{\odot}$ are impossible to reach the sensitive frequency scope of the LISA.

Our simulation indicate that compact IMBH X-ray binaries in most lifetime appear as normal X-ray sources, not ULXs. At present, compact IMBH X-ray binaries are rare in observations, which should arise from the selection effect. Comparing with stellar mass black hole, IMBH has a sole merit in forming compact binary systems. Because the loss rate of angular momentum by the gravitational radiation $\dot{J}_{\rm gr}\propto M_{\rm bh}^{5/2}$, $\dot{J}_{\rm gr}$ of IMBH (with a mass of $1000~M_{\odot}$) is five orders of magnitude higher than the stellar black hole (with a mass of $10~M_{\odot}$). Therefore, IMBH binaries can easily evolve into compact orbits, without invoking magnetic braking of Ap/Bp stars \citep{just06} or circumbinary disk \citep{chen06,chen15,chen19} like stellar-mass black-hole X-ray binaries. On the other hand, the characteristic strain of GW $h_{\rm c}\propto \mathcal{M}^{5/3}$, the chirp signals from IMBH X-ray binaries will be easily detected by the LISA. As a result, we propose that the detection of GW signal for some luminous X-ray sources with tight orbits could distinguish the nature of the accreting objects (IMBH or stellar-mass black hole).

Comparing with ultra-compact X-ray binaries, AM CVn, detached double white dwarf systems, the detectability that IMBH X-ray binaries will be detected by the LISA is not optimistic. First, IMBHs is very rare, and they may only exist in globular clusters and young dense clusters. Second, the rate that IMBHs accompany main sequence stars via via tidal or dynamical capture is very low. Third, the initial parameter space of compact IMBH X-ray binaries that LISA will detect is very narrow. Assuming each of 60 Galactic globular clusters hosts a $1000~M_{\odot} $ IMBH, the optimistic estimation for the number of IMBH X-ray binaries that LISA will detect in the Galaxy should be less than ten. Furthermore, it seems that the maximum detection distance for IMBH X-ray binaries can reach 1 Mpc, while the GW signals from double white dwarf systems are most likely dominant in the relevant area due to the initial mass function. Therefore, the extragalactic IMBH X-ray binaries are not expected to be detectable by the LISA. As a result, the detection probability of IMBH X-ray binaries by the LISA is not optimistic although they provide an opportunity of multi-messenger investigation. We expect that the observations of the LISA can confirm or rule out our idea in the future.

\acknowledgments {We acknowledge the anonymous referee for valuable comments
that have led to the improvement of the manuscript. We thank Xiang-Dong Li for useful
discussions. This work was partly supported by the National Natural Science Foundation of China (under grant Nos. 11573016, and 11733009), the Program for Innovative Research Team (in Science and Technology) at the
University of Henan Province.}

\end{document}